\documentclass{article}
\usepackage{hiph-art}
\volnumber{00} \issuenumber{1} \edyear{2005}                             
\frompage{000} \topage{000}                                              
\recrevdate{October 2005}                                              
\def\rightmark{Isotropization and Thermalization in Heavy Ion Collisions}
\def\leftmark{Yu. V. Kovchegov}
 
\title{{\bf Isotropization and Thermalization in Heavy Ion Collisions}} 
\authors{ 
{Yuri V. Kovchegov %
\index{Kovchegov, Yu.V.} 
}\\[2.812mm]
{\normalsize
Department of Physics, The Ohio State University \\ 
Columbus, OH 43210, USA\\[0.2ex] 
}}
 
\abstract{We argue that isotropization and, consequently, thermalization of the
system of gluons and quarks produced in an ultrarelativistic heavy ion
collision does not follow from Feynman diagram analysis to any order
in the coupling constant. We conclude that the apparent thermalization
of quarks and gluons, leading to success of perfect fluid
hydrodynamics in describing heavy ion collisions at RHIC, can only be
attributed to the non-perturbative QCD effects not captured by Feynman
diagrams. \\
\hspace*{4mm} We proceed by modeling these non-pertrubative thermalization 
effects using viscous hydrodynamics. We point out that matching Color
Glass Condensate inital conditions with viscous hydrodynamics leads to
continuous evolution of all components of energy-momentum tensor and,
unlike the case of ideal hydrodynamics, does not give a discontinuity
in the longitudinal pressure. An important consequence of such a
matching is a relationship between the thermalization time and shear
viscosity: we observe that small viscosity leads to short
thermalization time.}


 
\usepackage{epsfig}
\def\eq#1{{Eq.~(\ref{#1})}}
\def\fig#1{{Fig.~\ref{#1}}}
\newcommand{\be}{\begin{equation}}
\newcommand{\ee}{\end{equation}}
\newcommand{\ben}{\begin{eqnarray*}}
\newcommand{\een}{\end{eqnarray*}}
\newcommand{\as}{\alpha_s}
\newcommand{\un}{\underline}

\makeindex
\begin{document}
 
\maketitle

\section{Introduction: Isotropization Versus Free Streaming}

The results presented here are mainly based on the work done in
\cite{me1,me2}, but also include new developments described in Sect. 4.

Similar to the original Bjorken hydrodynamics approach \cite{bj}, let
us consider a central high energy collision of two very large
nuclei. For simplicity, here we will discuss the case where the
distribution of particles is independent of space-time rapidity $\eta
= (1/2) \ln (x_+ /x_-)$, where $x_\pm = (t \pm z)/\sqrt{2}$. Since the
nuclei are very large the transverse coordinate dependence can also be
neglected for most physical quantities, leaving only the dependence on
the proper time $\tau = \sqrt{2 x_+ x_-}$. For this geometry, one can
show that the most general energy-momentum tensor can be written as
(at $z=0$) \cite{me1}
\begin{eqnarray}\label{tmngen0}
 T^{\mu\nu} &=&  
 \left( \matrix{ \epsilon (\tau) & 0 & 0 & 0 \cr
  0 & p (\tau) & 0 & 0 \cr
  0 & 0 & p (\tau) & 0  \cr
  0 & 0 & 0  & p_3 (\tau) \cr} \right)\, ,
\end{eqnarray}
where the $z$-axis is taken along the beam direction, and $x,y$-axes
are in the transverse direction. Applying the conservation of
energy-momentum tensor condition
\be\label{cons}
\partial_\mu T^{\mu\nu} \, = \, 0
\ee
to the energy-momentum tensor that gives \eq{tmngen0} at $z=0$ we
obtain
\be\label{hydroeq}
\frac{d \epsilon}{d \tau} \, = \, - \frac{\epsilon + p_3}{\tau}.
\ee
There are two interesting cases one can consider:
\begin{itemize}
\item[(i)] if $p_3 =0$ longitudinal pressure vanishes and, due to 
\eq{hydroeq}, we get 
\be\label{efree}
\epsilon \sim \frac{1}{\tau}, 
\ee
such that the total energy $E \approx \epsilon \, \tau =$ const. This
case is known as {\sl free streaming}: the system expands freely
without loosing any energy.
\item[(ii)] if $p_3 = p$ the energy-momentum tensor in \eq{tmngen0} becomes 
{\sl isotropic}. This is the case of ideal Bjorken hydrodynamics
\cite{bj}. \eq{hydroeq} with $p_3 = p$ was derived in \cite{bj}. If combined 
with the ideal gas equation of state, $\epsilon = 3 \, p$, it gives
\be\label{43}
\epsilon \sim \frac{1}{\tau^{4/3}}
\ee
or, for other equations of state,
\be\label{ed}
\epsilon \sim \frac{1}{\tau^{1 + \Delta}} \hspace*{1cm} \mbox{with}  \hspace*{1cm} 
\Delta \, > \, 0. 
\ee
\end{itemize}

\eq{hydroeq} demonstrates that changes in the total energy $E \approx \epsilon \, \tau$ 
(or, equivalently, deviations from $\epsilon \sim 1/\tau$ scaling) are
due to work done by the longitudinal pressure $p_3$. The classical
initial conditions in the Color Glass Condensate (CGC) approach
\cite{KNV} yield the free streaming final state with $p_3 = 0$. A
thermalized quark-gluon plasma (QGP) is characterized by non-zero
$p_3$, leading to the energy density scaling as shown in
\eq{ed}. Therefore, below we will understand {\sl isotropization},
which is the necessary condition for {\sl thermalization}, as
dynamical generation of non-zero longitudinal pressure $p_3 \neq 0$,
or, equivalently, deviations from the scaling of \eq{efree} leading to
the scaling of \eq{ed}.

\section{Formal Argument}

An extensive search of the diagrams which would bring in the desired
deviations from the scaling of \eq{efree} carried out by the author
did not yield any positive results: while many diagrams have
contributions to $\epsilon$ scaling as shown in \eq{ed}, such terms
are {\sl always} subleading additive corrections to the leading (at
late times) terms scaling as shown in \eq{efree}. In fact one can
construct an argument \cite{me1} demonstrating that the leading
contribution to energy density from any-order diagrams scales as
$\epsilon \sim 1/\tau$. The argument is presented below.

\begin{figure}
\begin{center}
\epsfxsize=5cm
\leavevmode
\hbox{\epsffile{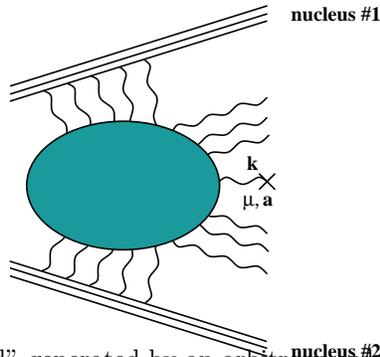}}
\end{center}
\vspace*{-1cm}
\caption{Gluon ``field'' generated by an arbitrary-order diagram (see text).}
\label{field}
\end{figure}

We begin by considering a gluon field generated by an arbitrary
Feynman diagram \cite{me1}, illustrated in \fig{field}. In
$\partial_\mu A^\mu = 0$ covariant gauge it can be written as
\be\label{ffi}
A_{\mu}^{a} (x) = - i \int \frac{ d^4 k}{(2 \pi)^4}\, \frac{e^{- i k
\cdot x} }{k^2 + i \epsilon k_0 } \, J_\mu^a (k),
\ee
where the function $J_\mu^a (k)$ denotes the rest of the diagram in
\fig{field} (the truncated part), which depends on the momenta of other 
outgoing gluons as well. Indeed gluon field can be defined as a simple
function only in the classical case: the ``field'' in \eq{ffi} should
be thought of as a Feynman diagram in \fig{field} with one of the
outgoing gluon lines being off mass-shell, i.e., a generalization of
the classical field which we will need in calculating energy density
\cite{me1}. (The expression in \eq{ffi} can also be thought of as an
operator equation.)

Substituting \eq{ffi} into the expression for energy-momentum tensor
\be
T^{\mu\nu} \, = \, \left< - F^{a \, \mu\rho} \, F^{a \, \nu}_{\ \ \
\rho} + \frac{1}{4} \, g^{\mu\nu} \, (F^a_{\rho\sigma})^2 \right>,
\ee
averaging over the nuclear wave functions and employing the symmetries
of the collision of two identical nuclei we obtain the energy density
due to the gluon field \cite{me1}
\ben\label{edens1}
\epsilon = \hspace*{-.6mm}\int \frac{ d^4 k \, d^4 k'}{(2 \pi)^8}
\frac{e^{-i k \cdot x - i k' \cdot x}}{(k^2 + i \epsilon k_0) (k'^2
+ i \epsilon k'_0)} 
\een
\be
\times \, \Bigg\{  \frac{1}{2} \left[ \left(
\frac{\tau}{x_+} \right)^2  \hspace*{-2mm} k_+ k'_+ -  {\un k}^2
\right] \hspace*{-1mm}f_1 (k^2, k'^2,  k \cdot k', k_T)  + \ldots \Bigg\}
\ee
where $f_1 (k^2, k'^2, k \cdot k', k_T)$ is some unknown function (a
``form-factor'') and the ellipsis indicate addition of two more
similar terms with different ``form-factors'' $f_2$ and
$f_3$. 

Rewriting each ``form-factor'' as
\ben
f_i (k^2 , k'^2 , k \cdot k', k_T) \, = \, f_i (k^2 =0, k'^2 =0, k
\cdot k' =0, k_T) + 
\een
\be\label{iter1}
+ [f_i (k^2 , k'^2 , k \cdot k', k_T) - f_i (k^2 =0, k'^2 =0, k \cdot k' =0,
k_T)]
\ee
and using the fact that the square of truncated part of the diagram
gives a cross section
\be
\frac{dN}{d^2 k \, dy} \, = \, \frac{1}{2 (2 \pi)^3} \, 
\bigg\langle\bigg\langle J^{a \, \rho} (k)
\, J^a_\rho (-k) \bigg\rangle\bigg\rangle \bigg|_{k^2 = 0}
\ee
we conclude that, keeping only the first term on the right hand side of
\eq{iter1} for all ``form-factors'' in \eq{edens1} yields
\be\label{edens2}
\epsilon \, = \, \frac{\pi}{2} \,  \int d^2 k \, 
\frac{d N}{d^2 k \, d \eta \, d^2 b} 
\ k_T^2 \, \left\{ \left[ J_1 (k_T \tau) \right]^2 + \left[ J_0 (k_T \tau) \right]^2 
\right\} \, \approx \, \frac{1}{\tau} \, \int d^2 k \, 
\frac{d N}{d^2 k \, d \eta \, d^2 b} \ k_T,
\ee
where the last equality is valid for late proper times $\tau$. Since,
as was shown in \cite{me1}, each factor of $k^2$, $k'^2$ or $k \cdot
k'$ gives a factor of $1/\tau$, the terms in the square brackets of
\eq{iter1} give a subleading (compared to \eq{edens2}) contribution 
to energy density at late times $\tau$ and can be safely
neglected.\footnote{We assume that $f_i$'s are continuous functions of
their arguments.} We have shown that any diagram and/or any set of
diagrams contributing to gluon production cross section lead to energy
density scaling as in \eq{efree}, i.e., that isotropization and,
consequently, thermalization do not take place in perturbation theory
analysis of the collisions.

The main assumption of the argument presented above is the existence
of multiplicity of produced gluons $dN/d^2 k \, dy$, which is the
essential assumption of QCD perturbation theory. This is what makes
our argument perturbative.

\section{Physical Argument}

Now let us present a physical argument demonstrating the origin of the
power of $4/3$ in \eq{43} and explaining why it is impossible to
achieve in perturbation theory \cite{me2}. Let us assume that
thermalization does take place at some time $\tau_{th}$. If a gauge
invariant time $\tau_{th}$ exists, we can put the QCD coupling
constant $g=0$ for all times $\tau > \tau_{th}$ without violating
gauge invariance. Bjorken hydrodynamics in the $g=0$ limit is governed
by the ideal gas equation of state $\epsilon = 3 \, p$, which leads to
the energy density scaling as shown in \eq{43}. (For small but
non-zero $g$, \eq{43} would get an $o (g^2)$ negative correction to
$4/3$: the expansion in $g$ would still be around the power of $4/3$.) 
Due to \eq{hydroeq}, the scaling of \eq{43} in the $g=0$ limit of
Bjorken hydrodynamics means that $p_3 \neq 0$ and the gas of {\sl
non-interacting} particles is {\sl doing work} in the longitudinal
direction! What causes such a behavior of the system? The problem lies
in the ideal gas equation of state, $\epsilon = 3 \, p$, which assumes
that the ideal gas is in contact with some {\sl external thermal
bath}. Such external thermal bath could be a background field or a box
containing the gas: the ideal gas of non-interacting particles stays
thermal through the interactions between the gas particles and the
thermal bath. This is the only interaction allowed in the $g
\rightarrow 0$ limit and it is responsible for the work done by the
non-interacting gas. Since there is no such external thermal bath in
heavy ion collisions, the scaling of \eq{43} is impossible to achieve
at small coupling.

Without the external thermal bath the particles in the gas would be
just free streaming, giving the physically correct energy density
scaling of \eq{efree}. Of course, at a fixed time $\tau$ hydro is not
applicable in the $g=0$ limit, since the mean free path of the
particles would exceed the longitudinal size of the system. However,
{\sl if} thermalization does happen, for any fixed arbitrary small
$g$, if we wait long enough hydrodynamics should become applicable,
leading to the scaling arbitrary close to that of \eq{43} and doing
work in the longitudinal direction which would be mostly due to
contact with the non-existing external thermal bath. Therefore, we
arrive at a contradiction, demonstrating that hydrodynamics is not
achievable at small coupling.  At large coupling, non-perturbative
effects may mimic the external thermal bath, possibly leading to
energy density scaling shown in \eq{ed}.

\section{Matching Color Glass Initial Conditions and Viscous Hydrodynamics}

If thermalization does take place via some non-perturbative mechanism,
understanding its dynamics would be extremely hard. Instead, let us
ask a pragmatic question: what is expected from a successful
thermalization scenario? Apart from giving us correct dynamical
mechanism for the generation of longitudinal pressure and the onset of
isotropization, one may expect a thermalization scenario to yield us
an estimate of the time when energy-momentum tensor would become
symmetric, $T_{\mu\nu} = \mbox{diag} \{ \epsilon, p, p, p \}$ at $z=0$
(again we are considering a central collision of two heavy ions in the
rapidity-independent approximation for produced particles). At this
time ideal hydrodynamics would be initiated and it could be used to
describe the subsequent evolution of the system. Unfortunately, as was
argued in \cite{KPSS}, ideal hydrodynamics is probably unachievable in
nature, since the shear viscosity $\eta$ never completely
vanishes.\footnote{Following the standard convention we will use
$\eta$ to denote shear viscosity in this chapter, which should not be
confused with space-time rapidity $\eta$ used in the previous
chapters.} Therefore, one has to use viscous hydrodynamics with the
energy-momentum tensor given by
\be\label{hydrovis1}
T_{\mu\nu} \, = \, (\epsilon + p) \, u_\mu \, u_\nu - p \, g_{\mu\nu}
+ \eta \, \left( \nabla_\mu u_\nu + \nabla_\nu u_\mu - \frac{2}{3} \,
\Delta_{\mu\nu} \, \nabla_\rho \, u^\rho \right),
\ee
where $u_\mu$ is the velocity profile, $\nabla_\mu \equiv (g_{\mu\nu}
- u_\mu \, u_\nu) \, \partial^\nu$ and $\Delta_{\mu\nu} \equiv
g_{\mu\nu} - u_\mu \, u_\nu$. In the boost-invariant case
\eq{hydrovis1} leads to
\begin{eqnarray}\label{tmnvis}
 T^{\mu\nu} &=& \left( \matrix{ \epsilon (\tau) & 0 & 0 & 0 \cr 0 & p
 (\tau) + \frac{2}{3} \, \frac{\eta}{\tau} & 0 & 0 \cr 0 & 0 & p
 (\tau) + \frac{2}{3} \, \frac{\eta}{\tau} & 0 \cr 0 & 0 & 0 & p
 (\tau) - \frac{4}{3} \, \frac{\eta}{\tau} \cr} \right)\, .
\end{eqnarray}
One can see that viscosity corrections increase transverse pressure
and decrease longitudinal pressure in the system. 

Now, if we follow \cite{KPSS} and assume that ideal hydrodynamics is
unachievable, we can try to address once again the question of what to
expect from a thermalization scenario. Our earlier answer would not
work anymore: thermalization can not give us a completely isotropic
energy-momentum tensor, since such a tensor is impossible due to
viscosity corrections. Therefore, thermalization dynamics can only
bring the system to a somewhat isotropic state achieved at certain
thermalization time, after which viscous hydrodynamics takes over.

However, the advantage of viscous hydrodynamics energy-momentum tensor
in \eq{tmnvis} is that it can be {\sl continuously} mapped onto the
energy-momentum tensor of Color Glass initial conditions without
including any additional thermalization dynamics. In the
rapidity-independent case the energy-momentum tensor due to the
classical fields in the Color Glass is given by \cite{KNV,me1}
\be\label{cgctmn}
T^{\mu\nu} = \mbox{diag} \{ \epsilon' (\tau), p' (\tau), p' (\tau), 0 \}.
\ee
We mark the energy density and pressure in \eq{cgctmn} with a prime to
distinguish them from the appropriate components of the
energy-momentum tensor in \eq{tmnvis}.

Indeed matching the energy-momentum tensors in Eqs. (\ref{cgctmn}) and
(\ref{tmnvis}) is a rather crude approximation, which, by omitting
thermalization dynamics, would only give us a lower bound on
thermalization time. Alternatively, one can view viscous hydrodynamics
(\ref{tmnvis}) as a model for the non-perturbative thermalization
effects: then, matching Eqs. (\ref{cgctmn}) and (\ref{tmnvis}) would
generate a non-perturbative thermalization scenario.

Requiring that at some ``thermalization'' time $\tau_0$ the components
of energy-momentum tensors in Eqs. (\ref{cgctmn}) and (\ref{tmnvis})
are equal yields the following set of equations:\footnote{Continuous
matching conditions in \eq{match} also lead to continuity of $d
\epsilon/d\tau$ due to \eq{hydroeq}.}
\be\label{match}
\left\{ \begin{array}{c} \epsilon (\tau_0) \, = \, \epsilon' (\tau_0) \\
p (\tau_0) + \frac{2}{3} \, \frac{\eta (\tau_0) }{\tau_0} \, = \, p'
(\tau_0) \\ p (\tau_0) - \frac{4}{3} \, \frac{\eta (\tau_0)}{\tau_0}
\, = \, 0. \end{array} \right.
\ee
 The energy density
$\epsilon'$ and transverse pressure $p'$ of the CGC initial conditions
determine, through Eqs.~(\ref{match}), the initial values of energy
density, pressure and viscosity of the QGP, along with the matching
time $\tau_0$. Indeed, the exact knowledge of QGP thermodynamics
would, in principle, allow one to express $\epsilon$, $p$ and $\eta$
as functions of the system's temperature $T$. Then Eqs.~(\ref{match})
would have only two unknowns -- matching time $\tau_0$ and the
corresponding temperature at the matching $T_0$, and would be
over-constrained. However, in the case when $T_\mu^\mu =0$ on both
sides of the matching, Eqs.~(\ref{match}) have only two independent
equations
\be\label{match2}
\left\{ \begin{array}{c} \epsilon (\tau_0) \, \approx \, \epsilon' (\tau_0) \\
p (\tau_0) \, \approx \, \frac{4}{3} \, \frac{\eta
(\tau_0)}{\tau_0} \end{array} \right.
\ee
and are not over-constrained anymore. The tracelessness condition,
$T_\mu^\mu =0$, is valid in CGC both at the classical level, and at
the level of leading logarithmic small-$x$ evolution. On QGP side
$T_\mu^\mu =0$ is valid at very high temperatures, and also $\epsilon
= 3 \, p$ appears to be a good approximation for the equation of state
in the strong coupling regime. We conclude that solution of
Eqs.~(\ref{match2}) in terms of $\tau_0$ and $T_0$ is likely to
satisfy the third equation in (\ref{match}) as well.

It is impossible to solve Eqs.~(\ref{match2}) without knowing the
exact dependence of $\epsilon$ and $p$ on temperature. Instead, we
will find an approximate solution by writing
\be\label{eos}
\epsilon (T) \, = \, 3 \, p (T) \, = \, \frac{\pi^2}{30} \, n (g) \, T^4, 
\ee
where $n (g)$ is some function of QCD coupling constant $g (T)$,
which, at high temperatures, corresponding to small couplings, just
counts the number of quark and gluon degrees of freedom. Following
\cite{KPSS} we write for the shear viscosity
\be\label{visc}
\eta \, = \, f (g^2 \, N_c) \, N_c^2 \, T^3
\ee
with $f (g^2 \, N_c)$ some function of the coupling. For the Color
Glass energy density we write
\be\label{ecgc}
\epsilon' \, = \, c_E \, \frac{C_F \, Q_s^3}{\as \, \tau \, 2 \pi^2},
\ee
where $c_E$ is some order $1$ coefficient to be determined by explicit
calculations and $Q_s$ is the saturation scale. Substituting
Eqs. (\ref{eos}), (\ref{visc}) and (\ref{ecgc}) into
Eqs.~(\ref{match2}), and neglecting, perhaps unreasonably, the
dependence of the coupling constant $g(T)$ on the temperature $T$, we
obtain the matching time
\be\label{time}
\tau_0 \, \approx \, \frac{\as^{1/3} \, [f (g^2 \, N_c) \, N_c^2]^{4/3}}{Q_s \, n (g)} 
\, \frac{120}{\pi^2} \, \left( \frac{8 \, \pi^2}{c_E \, C_F} \right)^{1/3},
\ee
which is a good estimate of thermalization time being the only
relevant time scale.

Let us analyze the thermalization time from \eq{time}. First of all,
while, as we argued in the previous Sections, perturbative
thermalization is dynamically impossible, we can still explore the
small-coupling asymptotics of \eq{time} to construct a {\sl lower}
bound on perturbative thermalization time. To that end we note that at
small values of the coupling $n (g)$ approaches a constant, while $f
(g^2 \, N_c) \, \sim \, 1/g^4$. Using this in \eq{time}, and dropping
all the constant factors, gives
\be
\tau_0 \, \ge \, \frac{1}{\as^{7/3} \, Q_s} ,
\ee
which is precisely the lower bound advocated by Arnold and Lenaghan in
\cite{AL}.

In general, using \eq{visc} in \eq{time} to replace $f (g^2 \, N_c) \,
N_c^2$ with $\eta/T^3$ and dropping the numerical factors we conclude
that
\be\label{centr}
\tau_0 \, \propto \, \frac{\as^{1/3}}{Q_s} \, \left( \frac{\eta}{T^3} \right)^{4/3}.
\ee
This is the central result of this Section. It appears that the
matching procedure between the CGC initial conditions and viscous
hydrodynamics leads to dependence of thermalization time on shear
viscosity. Moreover, the dependence shown in \eq{centr} implies that
{\sl lower viscosity leads to shorter equilibration time}! This
conclusion agrees well with the fact that hydrodynamic simulations
demand small shear viscosity and early thermalization time to describe
RHIC data. It appears that, if our matching procedure captures
correctly at least some features of the actual non-perturbative
thermalization dynamics, short thermalization time and small viscosity
of the quark-gluon plasma may be related.

Finally, assuming that 't Hooft coupling $g^2 N_c$ is large at RHIC,
one can use the result of \cite{KPSS} that $f (g^2 \, N_c) = \pi /8$
in the strong coupling limit (indeed, of ${\cal N}=4$ SYM theory), to
estimate the thermalization time $\tau_0$ using \eq{time}. Using $\as
= 0.3$, $c_E = 1$, $Q_s = 1.4$~GeV for central RHIC collisions, and
$n(g) = 47.5$ for three flavors yields $\tau_0 \approx 0.5$~fm, in a
very good agreement with thermalization time required by hydrodynamic
simulations at RHIC.

\section*{Acknowledgments}

I would like to thank Ulrich Heinz, Larry McLerran and Al Mueller for
many informative discussions. 

This work is supported in part by the U.S. Department of Energy under
Grant No. DE-FG02-05ER41377.

\end{document}